# Quaternionic Quantum Mechanics: the Particles, their q-Potentials and Mathematical Electron Model

Bogusław Bożek, Marek Danielewski and Lucjan Sapa*

AGH UST, Mickiewicza 30, 30-059 Kraków, Poland; sapa@agh.edu.pl

**Abstract.** In this work we show the quaternionic quantum descriptions of physical processes from the Planck to macro scale. The results presented here are based on the concepts of the Cauchy continuum and the elementary cell at the Planck scale. The symmetrization of quaternion relations and the postulate of quaternion velocity have been crucial in the present development. The momentum of the expansion and compression $\dot{u}_0(t,x)$ is the consequence of the scalar term $\sigma_0(t,x)$ in the quaternionic deformation potential. The key new results are listed below:

- The quaternionic $G_0(m)(\sigma_0 + \hat{\phi})$, vectorial $G_0(m)\hat{\phi}$ and scalar $G_0(m)\sigma_0$ propagators are used to generate the second order PDE systems for the proton, electron and neutron.
- The mathematical model of an electron is formulated. It is described by the hyperbolic-elliptic partial differential system of quaternion equations with the initial-boundary conditions. The boundary conditions are generated by the quaternion energy flux that is found with the use of the Gauss theorem, the Cauchy-Riemann derivative and other mathematical formulas.
- The rigorous assessment of the second order PDE systems allows proposing the two second order PDE systems for the *u* and *d* quarks from the *up* and *down* groups.
- It was verified that both the proton and the neutron obey experimental findings and are formed by three quarks. The proton and neutron are formed by the *d-u-u* and *d-d-u* complexes, respectively.
- The *u* and *d* quarks do not comply with the Cauchy equation of motion. The inconsistencies of the quarks' PDE with the quaternion forms of the Cauchy equation of motion account for their short lifetime and the observed Quarks Chains. That is, explain the Wilczek phenomenological paradox: *"Quarks are Born Free, but everywhere they are in Chains"*.

**Keywords:** electron, quaternionic potential, vectorial potential, proton, electron, quark chains

# 1. Introduction

After developing the Hilbert space formulation of quantum mechanics [1], von Neumann looked into higher mathematical analysis, i.e., rings of operators, in an attempt to get rid of some of the *ad hoc* postulates of quantum mechanics. In 1936 with Birkhoff suggested the Quaternionic Quantum Mechanics (QQM), where wave functions or probability amplitudes are quaternion valued [2]. But systematic work on the quaternionic extension of quantum mechanics has not begun. The essential results relevant to the present paper are by Lanczos. His dissertation was on a quaternionic field theory of the classical electrodynamics [3,4]. In his derivation of the Dirac equation [5], there is a doubling in the number of solutions and the concepts that still remain at the front of the fundamental theory. These articles were unnoticed by contemporaries; Lanczos abandoned quaternions and never returned to Quaternionic Field Theory (QFT).

Almost immediately it was demonstrated that the Cauchy-Riemann type conditions in the quaternion representation are identical in the shape to vacuum equations of electrodynamics [6] and that the Dirac transition amplitudes are quaternion valued [7]. Christianto derived an original wave equation from the correspondence between the Dirac equation and the Maxwell electromagnetic equations via the biquaternionic representation [8]. The Adler schema of the quaternionizing the quantum mechanics inspired the Harari-Shupe's model for the composite quarks and leptons [9,10] and the substantial progress in the QQM and QFT [11]. Adler presented a major conceptual advance for the purpose of determining whether a quaternionic Hilbert space is the suitable for the unification of the standard model forces with gravitation. He provided an introduction to the problem of formulating quantum field theories and concluded that the QQM may fit into the physics of unification and measurement theory issues [12].

The focus here is on quaternion quantum mechanics and quaternionic field theory. The QQM presented here is ontological in a sense that it starts with being, that are the Cauchy ideal elastic continuum at the macro-scale ($> 10^{-20}$ m) and the "Planck unit cell" at the microscale ($\sim 10^{-35}$ m) [15,13]. The basic categories of being and their relations are governed by the quaternion algebra [15].

The evolution of the Planck-Kleinert Crystal (P-KC) model and the development of the QQM are shown in succeeding papers [14,15,16,17,18]. In this article we present the QQM in the most recent, refined form:

- We use the ontology-based formalism which is based on the Planck-Kleinert crystal concept [13] and the quaternion algebra introduced by Hamilton, section 1.1.
- The widely used the Helmholtz decomposition is used in the general form in $\mathrm{R}^4$, section 2.
- The all vectors are in the $\mathrm{R}^4$ representation, e.g., the four-velocity is the "new" variable that allows for the symmetrization of the Hamiltonian [19] and the first and second order wave equations.
- The second order PDE systems of the quark particles are proposed.
- The mathematical model of an electron is formulated. It is described by the hyperbolic-elliptic partial differential system of quaternion equations with the initial-boundary conditions. This differential problem is decomposed onto the hyperbolic equation with the Neumann boundary condition on compression and the hyperbolic-elliptic subsystem on twist with some specific boundary condition concerning rotation. The boundary conditions are generated by the quaternion energy flux that is found with the use of the Gauss theorem, the Cauchy-Riemann derivative and other mathematical formulas.
- The further studies in order to verify or refute those propositions are suggested.

The Cauchy model of the elastic continuum is presented in Section 1.2. We construct a Lagrangian with the use of the Cauchy–Riemann operator and introduce the key new concept, the quaternion valued velocity, section 2. Abbreviations used in the text are presented in Appendix A.

## 1.1. Quaternions

The elements of the quaternion algebra used in the QQM and QFT were already presented in [14-16]. The basic definitions and formulas of quaternions and quaternionic functions can be found in [20,21].

In Hamilton's own words, he created the $R^4$ analog of complex numbers as the equivalent of the time-space continuum [22]:

*"Time is said to have only one dimension, and space to have three dimensions. The mathematical quaternion partakes of both these elements; in technical language it may be said to be 'time plus space'... and in this sense it has, or at least involves a reference to, four dimensions."*

We demonstrate here that, the Hamilton's *'time plus space'* is consistent with the Cauchy model of ideal elastic continuum in the quaternionic representation.

The algebra of quaternions $q = q_0 + \hat{q} \in \mathbb{Q}$ has all properties of an algebra with the unity.

1. The quaternionic deformation potential, i.e., the deformation four-potential or q-potential is a relativistic function from which the displacement field can be derived. It combines both a compression scalar potential $q_0$ and a torsion vector potential $\hat{q}$ (twist) into a single quaternion (four-vector) $q = q_0 + \hat{q} \in \mathbb{Q}$.
2. The multiplication of quaternions is noncommutative: $p \cdot q \neq q \cdot p$. The noncommutativity of multiplication is the only property that makes quaternions different from real and complex numbers.
3. The quaternionic deformation potential is invariant in the sense of Lorentz.

An arbitrary quaternion $q \in \mathbb{Q}$ can be written in the algebraic form

$$q = q_0 + q_1 i + q_2 j + q_3 k, \tag{1}$$

where $i, j, k$ are called the imaginary units and fulfill the relations:

$$i^2 = j^2 = k^2 = -1, \quad ij = -ji = k, \quad jk = -kj = i, \quad ki = -ik = j. \tag{2}$$

There is possible also to represent quaternions by some matrices.

The commutator of two elements $p$ and $q$ is defined by the formula

$$[p, q] = p \cdot q - q \cdot p = 2\hat{p} \times \hat{q} \tag{3}$$

and can be looked at as a measure of noncommutativity. Two quaternions commute, i.e., $[p, q] = 0$ if and only if their vector parts $\hat{p}$ and $\hat{q}$ are collinear.

Quaternion-valued functions describe a lot of useful physical models, e.g., the electric and magnetic fields [23], section 3. Let $\Omega \subset R^3$ be a bounded set. The so-called $\mathbb{Q}$-valued functions have the form

$$q(x) = q_0(x) + q_1(x)i + q_2(x)j + q_3(x)k, \, x = (x_1, x_2, x_3) \in \Omega, \tag{4}$$

where the functions $q_0(x), q_l(x), \, l = 1, 2, 3$ are real-valued. Similarly, the functions $q(t, x)$, depending on time $t$, may be considered. Properties such as continuity, differentiability, integrability, and so on, have to be possessed by all the components $q_0(t, x), q_l(t, x), \, l = 1, 2, 3$. In this manner, the Banach, Hilbert, and Sobolev spaces of $\mathbb{Q}$-valued functions can be defined [23]. In the Hilbert space over $\mathbb{Q}$

$$L^2(\Omega) = \left\{ q : \Omega \to \mathbb{Q} \,\middle|\, \int_\Omega q_0^2 \, dx < \infty, \int_\Omega q_l^2 \, dx < \infty, l = 1, 2, 3 \right\}, \tag{5}$$

the inner product can be defined as follows

$$\langle q_1, q_2 \rangle = \int_\Omega q_1 \cdot q_2^* \, dx, \quad q_1, q_2 \in L^2(\Omega). \tag{6}$$

The Fourier series, Lebesgue measure, Gelfand triples, Laplace transform, and many others on the vector space of $\mathbb{Q}$-valued functions over $\mathbb{Q}$ can be defined in a standard way as in the real and complex cases.

Moreover, very useful in applications is the exponential quaternionic function of quaternionic variable that has trigonometrical representation

$$e^q = e^{q_0} \left( \cos|\hat{q}| + \hat{q}/|\hat{q}| \sin|\hat{q}| \right), \quad q = q_0 + \hat{q} \in \mathbb{Q}. \tag{7}$$

In the further part we use the Cauchy–Riemann operator $D$ acting on the quaternion-valued functions $q = q_0 + \hat{q}$ of the form

$$Dq = -\operatorname{div}\hat{q} + \operatorname{grad} q_0 + \operatorname{rot}\hat{q}. \tag{8}$$

**Remark 1.** *Hurwitz's theorem says that there are only four normed division algebras: $\mathbb{R}, \mathbb{C}, \mathbb{Q}$ and the octonions algebra. Lagrange's four-square theorem in number theory states that every non-negative integer is the sum of four integer squares. This theorem may have applications in QQM.*

### 1.2. The Cauchy Dispacement Field: the Theory of Elasticity and the Properties at the Planck Scale

Cauchy finished the theory of the ideal elastic continuum in 1822 [24], right away Poisson [25] studied the elementary waves. Neumann [26] gave the proof of uniqueness of solutions of some initial-boundary value problems. The rigorous completeness proof was given by Duhem [27]. The Cauchy theory is the first real, well posed theory of elasticity using the continuum approach, where the macroscopic phenomena are described in the terms of field variables [28]: the compression $\operatorname{div}\mathbf{u}$, and the twist $\operatorname{rot}\mathbf{u}$. The stress tensor of the ideal elastic continuum is given by

$$\mathbf{T} = \lambda \operatorname{tr}(\mathbf{D})\mathbf{I} + 2\mu\mathbf{D}, \tag{9}$$

where $\operatorname{tr}(\mathbf{D})$ is the trace of the strain tensor, $\mathbf{I}$ is the identity matrix and the two moduli of elasticity, $\lambda$ and $\mu$, are the material-dependent constants. It was shown by Cauchy and Saint Venant that if the particles composing a regular crystal interact pairwise through central forces, then there is an additional symmetry that implies the Poisson ratio 0.25 and equal Lamé's coefficients: $\lambda = \mu$ [29]. The identity $\operatorname{grad}\operatorname{div}\mathbf{u} = \operatorname{div}\operatorname{grad}\mathbf{u} + \operatorname{rot}\operatorname{rot}\mathbf{u}$ implies that the stress tensor becomes:

$$\begin{aligned}\operatorname{div}\mathbf{T} &= 2\mu\operatorname{grad}\operatorname{div}\mathbf{u} + \mu\operatorname{div}\operatorname{grad}\mathbf{u} \\ &= 3\mu\operatorname{grad}\operatorname{div}\mathbf{u} - \mu\operatorname{rot}\operatorname{rot}\mathbf{u}.\end{aligned} \tag{10}$$

The Cauchy equation of motion generalizes: (1) the Newton laws of motion (the conservation of the linear and angular momenta) to an ideal elastic solid, and (2) the concept of stress in terms of the gradients in the displacement field $\mathbf{u}(t,x) \in \mathbb{R}^3$:

$$\rho\frac{\partial^2 \mathbf{u}}{\partial t^2} = 3\mu\operatorname{grad}\operatorname{div}\mathbf{u} - \mu\operatorname{rot}\operatorname{rot}\mathbf{u} + \mathbf{F}, \tag{11}$$

where $\mathbf{F}$ is the force induced solely by the displacements caused by the entities already present in the elastic continuum. In the next sections we show that the force must be generalized and follows from the Cauchy-Riemann derivative of the deformation potential $f$ caused by the particle waves, i.e., the force in $\mathbb{R}^4$ is the consequence of the 4-potential $f$ produced by the particles.

From Equation (11) the vectorial representation of the energy density in the deformation field can be computed [28,30]

$$e = \frac{1}{2}\dot{\mathbf{u}}\circ\dot{\mathbf{u}} + \frac{3}{2}c^2\left(\operatorname{div}\mathbf{u}\right)^2 + \frac{1}{2}c^2\operatorname{rot}\mathbf{u}\circ\operatorname{rot}\mathbf{u}, \tag{12}$$

where $\dot{\mathbf{u}} = \partial\mathbf{u}/\partial t$.

In the following we consider the Cauchy continuum with Face Centered Cubic (FCC) structure. The Young modulus $Y$ describes tensile elasticity which is axial stiffness of the length of a body to deformation along the axis of the applied tensile force. It is related to Lamé's two moduli of elasticity by

$$Y = \mu(3\lambda + 2\mu)/(\lambda + \mu) \stackrel{\lambda=\mu}{=} 2.5\mu. \qquad (13)$$

If $l_P$ denotes the dimension of the FCC elementary cell that consists of the four interacting Planck particles showing the mass $m_P$, the Planck density equals: $\rho_P = 4m_P/l_P^3 = const$. The computed Planck density, the Young's modulus and, the other properties of the Cauchy continuum at the Planck scale are shown in Table 1.

**Table 1.** The physical constants of the Cauchy continuum (fcc ideal isotropic crystal).

| Label Used in This Work | Planck Constants | Symbol for Unit | Value | SI Unit | Reference |
|---|---|---|---|---|---|
| Lattice parameter | Planck length | $l_P$ | $1.616229(38) \times 10^{-35}$ | M | [33] |
| Poisson ratio | | $\nu$ | 0.25 | - | [29] |
| Mass of the Planck particle | Planck mass | $m_P$ | $2.176470(51) \times 10^{-8}$ | kg | [33] |
| Duration of the internal process | Planck time | $t_P$ | $5.39116(13) \times 10^{-44}$ | $s^{-1}$ | [33] |
| Transverse wave velocity | Light velocity in vacuum | $c$ | $2.99792458 \times 10^{8}$ | m·s$^{-1}$ | $c = l_P/t_P$ [33] |
| Planck density | | $\rho_P$ | $2.062072 \times 10^{97}$ | kg·m$^{-3}$ | [33] |
| Young's modulus Intrinsic energy density | | $Y$ | $4.6332447 \times 10^{114}$ | kg·m$^{-1}$s$^{-2}$ | $Y = 2.5\rho_P c^2$ |

We consider the small deformation limit and the negligible density changes. It allows assuming the constant transverse wave velocity: $c = \sqrt{\mu/\rho_P} = const.$ [28] and Equation (11) becomes:

$$\frac{1}{c^2}\frac{\partial^2 \mathbf{u}}{\partial t^2} = 3\,\text{grad div}\,\mathbf{u} - \text{rot rot}\,\mathbf{u} + \frac{1}{\mu}\mathbf{F}. \qquad (14)$$

The Cauchy and to the same degree the majority of physical problems cannot be reduced to vectorial models (the vector product does not permit the formulation of algebra with unity, for example, the division operation is not defined). By acting on the equation (14) by rot and div operators we separate the transverse and the longitudinal processes:

$$\begin{aligned}\text{div}\left(\frac{1}{c^2}\frac{\partial^2 \mathbf{u}}{\partial t^2} = 3\nabla(\nabla\cdot\mathbf{u}) - \nabla\times(\nabla\times\mathbf{u}) + \frac{1}{\mu}\mathbf{F}\right) \\ \text{rot}\left(\frac{1}{c^2}\frac{\partial^2 \mathbf{u}}{\partial t^2} = 3\nabla(\nabla\cdot\mathbf{u}) - \nabla\times(\nabla\times\mathbf{u}) + \frac{1}{\mu}\mathbf{F}\right)\end{aligned} \xRightarrow{\nabla\times(\nabla\times\mathbf{u}_\phi)=-\Delta\mathbf{u}_\phi} \begin{cases}\frac{1}{c^2}\frac{\partial^2}{\partial t^2}\text{div}\,\mathbf{u}_0 = 3\Delta\,\text{div}\,\mathbf{u}_0 + \frac{1}{\mu}\text{div}\,\mathbf{F} \\ \frac{1}{c^2}\frac{\partial^2}{\partial t^2}\text{rot}\,\mathbf{u}_\phi = \Delta\,\text{rot}\,\mathbf{u}_\phi + \frac{1}{\mu}\text{rot}\,\mathbf{F}\end{cases}. \qquad (15)$$

The Cauchy equation of motion combined with the Helmholtz decomposition theorem results in four second-order scalar differential equations, "quattro cluster", and implies the transverse and longitudinal waves in the Cauchy elastic solid.

**Remark 2**
1. *The mathematical analysis confirms that the Cauchy model is well posed, i.e., it has a unique solution and its behavior changes continuously with respect to the initial conditions* [27].

2. *The Helmholtz decomposition is never unique* [31].
3. *The Hamilton algebra of quaternions and its relation to the four-dimensional space allow combining the Cauchy theory with the electrodynamics, gravity and quantum mechanics.*

## 2. Theory

**2.1 The Cauchy Deformation Field in the Quaternion Representation**

The Cauchy classical theory of elasticity is an elegant starting point to show the physical reality and the significance and beauty of quaternions. The Hamilton algebra $\mathbb{Q}$ allows recoupling the compression and twist that are separated in (15). Upon denoting $\sigma_0 = \text{div}\,\mathbf{u}_0 = (\sigma_0, 0, 0, 0)$ and $\hat{\phi} = \text{rot}\,\mathbf{u}_\phi = (0, \phi_1, \phi_2, \phi_3)$ we get

$$\begin{cases} \dfrac{\partial^2 \sigma_0}{\partial t^2} = 3c^2 \Delta \sigma_0 \\ \dfrac{\partial^2 \hat{\phi}}{\partial t^2} = c^2 \Delta \hat{\phi} \end{cases} \Leftrightarrow \left(\dfrac{\partial^2}{\partial t^2} - c^2 \Delta\right)\sigma - 2c^2 \Delta \sigma_0 = 0. \tag{16}$$

The decomposition $\mathbf{u} = \mathbf{u}_0 + \mathbf{u}_\phi$ in Equation (14) results in four equations in Equation (15) and implies the existence of the deformation field $\sigma = \sigma_0 + \hat{\phi}$ that represents the twist and compression fields as a superposition of real (scalar compression $\sigma_0$) and imaginary (twist vector $\hat{\phi}$) field parts at each point

$$\sigma = \sigma_0 + \hat{\phi} \in \mathbb{Q} \quad \text{and} \quad \sigma^* = \sigma_0 - \hat{\phi} \in \mathbb{Q}. \tag{17}$$

Adding equations in (16) one gets the quaternion form of the Cauchy motion equation

$$\frac{1}{c^2} \frac{\partial^2 \sigma}{\partial t^2} - \Delta \sigma - 2\Delta \sigma_0 = 0. \tag{18}$$

Since $\dot{\mathbf{u}} \circ \dot{\mathbf{u}} = \hat{\dot{u}} \circ \hat{\dot{u}} = -\hat{\dot{u}} \cdot \hat{\dot{u}} = \hat{\dot{u}} \cdot \hat{\dot{u}}^*$, where $\hat{\dot{u}} = \dot{u}_1 i + \dot{u}_2 j + \dot{u}_3 k$ and $\dot{\mathbf{u}} = (\dot{u}_1, \dot{u}_2, \dot{u}_3)$, the overall energy of the deformation field, the relation (14) takes the form

$$e = \tfrac{1}{2}\hat{\dot{u}} \cdot \hat{\dot{u}}^* + \tfrac{1}{2}c^2\, \sigma \cdot \sigma^* + c^2 \sigma_0^2, \tag{19}$$

where the velocity within the particle wave is given by the Cauchy-Riemann derivative

$$\hat{\dot{u}} = -\frac{\hbar}{m} D\sigma, \tag{20}$$

where $D\sigma = \text{grad}\,\sigma_0 + \text{rot}\,\hat{\phi}$, because $\text{div}\,\hat{\phi} = \text{div}\,\text{rot}\,\mathbf{u}_\phi = 0$. The overall energy in arbitrary volume $\Omega$ follows from Eq. (19):

$$E = \int_\Omega \rho_E(t, x)\,\mathrm{d}x = \int_\Omega \rho_P \left(\tfrac{1}{2}\hat{\dot{u}} \cdot \hat{\dot{u}}^* + \tfrac{1}{2}c^2\, \sigma \cdot \sigma^* + c^2 \sigma_0^2\right) \mathrm{d}x = mc^2, \tag{21}$$

where the external potential, e.g., $V(x)$, is omitted.

The kinetic energy in (19) follows from the vectorial form: $\tfrac{1}{2}\hat{\dot{u}} \cdot \hat{\dot{u}}^*$, that can be regarded as $\mathbb{R}^3$ representation that does not describe the volume changes. Contrary, the deformations have quaternion representation in $\mathbb{R}^4$. In (17) the quaternion potential, i.e., the deformation four-potential, is defined by

$$\begin{aligned}
\sigma &= \sigma_0 + \hat{\phi} \\
\begin{bmatrix} q\text{-potential,} \\ \text{deformation} \end{bmatrix} &= \begin{bmatrix} \operatorname{div} \mathbf{u}_0 \\ \text{compression} \end{bmatrix} + \begin{bmatrix} \operatorname{rot} \mathbf{u}_\phi \\ \text{twist} \end{bmatrix}.
\end{aligned} \qquad (22)$$

The quaternionic velocity $\dot{u} \in \mathbb{Q}$ should represent now all the deformation velocities in $\mathbb{R}^4$. In the electron model we demonstrate the practical application with an example of the particle wave showing the equivalent mass $m$:

$$\begin{aligned}
\dot{u} &= \dot{u}_0 + \hat{\dot{u}}, \\
-\frac{\hbar}{m} \mathrm{D}\sigma &= -\frac{\hbar}{m} \frac{\sigma_0}{l_P} - \frac{\hbar}{m}\left(\operatorname{grad} \sigma_0 + \operatorname{rot} \hat{\phi}\right). \\
\begin{bmatrix} \text{velocity of the} \\ q\text{-potential changes,} \\ \text{deformation velocity} \end{bmatrix} &= \begin{bmatrix} \text{compression} \\ \text{velocity} \end{bmatrix} + \begin{bmatrix} \text{twist velocity} \end{bmatrix}
\end{aligned} \qquad (23)$$

## 2.2. The boundary conditions

In order to obtain boundary conditions we will find the flux $S$ of the energy $e$. Let $V \subset \Omega$ be any subregion of $\Omega$ with the smooth boundary $\partial V$. The rate of change of the total energy within $V$ equals the negative of the net flux through $\partial V$:

$$\frac{\mathrm{d}}{\mathrm{d}t} \int_V e \, \mathrm{d}^3 \mathbf{r} = -\int_{\partial V} S \circ \mathbf{n} \, \mathrm{d}(\partial V), \qquad (24)$$

where $\mathbf{n} = (n_1, n_2, n_3) \in \mathbb{R}^3$ is the outside normal unit vector to $\partial V$. It follows from the Gauss theorem that

$$\int_V \frac{\partial e}{\partial t} \mathrm{d}^3 \mathbf{r} = -\int_V \operatorname{div} S \, \mathrm{d}^3 \mathbf{r}. \qquad (25)$$

Thus

$$\frac{\partial e}{\partial t} = -\operatorname{div} S \qquad (26)$$

as $V$ was arbitrary.

On the other hand the formula (12) upon differentiation becomes

$$\frac{\partial e}{\partial t} = \dot{\mathbf{u}} \circ \ddot{\mathbf{u}} + 3c^2 \operatorname{div} \mathbf{u} \operatorname{div} \dot{\mathbf{u}} + c^2 \operatorname{rot} \mathbf{u} \circ \operatorname{rot} \dot{\mathbf{u}} \qquad (27)$$

and using Eq. (14) with $\mathbf{F} = 0$ we have

$$\frac{\partial e}{\partial t} = \dot{\mathbf{u}} \circ \left(3c^2 \operatorname{grad} \operatorname{div} \mathbf{u} - c^2 \operatorname{rot} \operatorname{rot} \mathbf{u}\right) + 3c^2 \operatorname{div} \mathbf{u} \operatorname{div} \dot{\mathbf{u}} + c^2 \operatorname{rot} \mathbf{u} \circ \operatorname{rot} \dot{\mathbf{u}}. \qquad (28)$$

Taking into account the identities $\operatorname{div}(a\mathbf{u}) = \mathbf{u} \circ \operatorname{grad} a + a \operatorname{div} \operatorname{grad} \mathbf{u}$ and $\operatorname{div}(\mathbf{A} \times \mathbf{B}) = \mathbf{B} \circ \operatorname{rot} \mathbf{A} - \mathbf{A} \circ \operatorname{rot} \mathbf{B}$ the formula (28) becomes

$$\frac{\partial e}{\partial t} = -\operatorname{div}\left[c^2 (\operatorname{rot} \mathbf{u}) \times \dot{\mathbf{u}} - 3c^2 (\operatorname{div} \mathbf{u}) \dot{\mathbf{u}}\right]. \qquad (29)$$

Comparing (26) and (29), the energy flux equals $S = c^2 (\text{rot}\,\mathbf{u}) \times \dot{\mathbf{u}} - 3c^2 (\text{div}\,\mathbf{u}) \dot{\mathbf{u}}$ or in the quaternionic notation

$$\hat{S} = c^2 \hat{\phi} \times \hat{\dot{u}} - 3c^2 \sigma_0 \hat{\dot{u}}, \tag{30}$$

or equivalently

$$\hat{S} = c^2 (\sigma - \sigma_0) \times \hat{\dot{u}} - 3c^2 \sigma_0 \hat{\dot{u}}. \tag{31}$$

Thus, the relation (29) can be written in the form of the quaternionic continuity equation

$$\frac{\partial e}{\partial t} + \text{div}\,\hat{S} = 0. \tag{32}$$

Because nothing flows over the boundary $\partial V$, we assume that

$$\hat{S} \circ \hat{n} = 0 \tag{33}$$

on $\partial V$ where $\hat{n} = n_1 i + n_2 j + n_3 k$. Equation (33) will be used in the construction of the boundary conditions of an electron model.

### 2.3. Maxwell Equations in the Cauchy Continuum

We consider system (15) with the nonzero external force field $\mathbf{F}$

$$\begin{cases} \dfrac{1}{c^2} \dfrac{\partial^2}{\partial t^2} \text{div}\,\mathbf{u}_0 - 3\Delta \text{div}\,\mathbf{u}_0 = \dfrac{1}{\mu} \text{div}\,\mathbf{F} \\ \dfrac{1}{c^2} \dfrac{\partial^2}{\partial t^2} \text{rot}\,\mathbf{u}_\phi - \Delta \text{rot}\,\mathbf{u}_\phi = \dfrac{1}{\mu} \text{rot}\,\mathbf{F} \end{cases}. \tag{34}$$

The force $\mathbf{F}$ is due to the deformations $\text{div}\,\mathbf{u}$ and $\text{rot}\,\mathbf{u}$ caused by the presence of the flux of the external charged particles, i.e., there exists the external displacement field

$$f = f_0 + \hat{f} \in \mathbb{Q}. \tag{35}$$

The force field follows from the Cauchy-Riemann derivative and the condition $\text{div}\,\hat{f} = 0$

$$\mathbf{F} = -\mu D f = -\mu \,\text{grad}\, f_0 - \mu \,\text{rot}\, \hat{f}. \tag{36}$$

Combining (34), (36) and the definitions in (22), the system (34) becomes

$$\begin{cases} \dfrac{1}{c^2} \dfrac{\partial^2 \sigma_0}{\partial t^2} - 3\Delta \sigma_0 = -\Delta f_0 \\ \dfrac{1}{c^2} \dfrac{\partial^2 \hat{\phi}}{\partial t^2} - \Delta \hat{\phi} = \Delta \hat{f} \end{cases}. \tag{37}$$

By noting that the negative $f_0$ is a result of net inflow of charged particles and it results in the positive change of $\sigma_0$ we get: $\Delta f_0 - \Delta \sigma_0 = 0$, the scalar equation in (37) becomes symmetric

$$\begin{cases} \dfrac{1}{c^2}\dfrac{\partial^2 \sigma_0}{\partial t^2} - \Delta \sigma_0 = \Delta f_0 \\ \dfrac{1}{c^2}\dfrac{\partial^2 \hat{\phi}}{\partial t^2} - \Delta \hat{\phi} = \Delta \hat{f} \end{cases}. \tag{38}$$

Upon adding equations in (38) one gets quaternionic representation of the Maxwell displacements

$$\left(\dfrac{1}{c^2}\dfrac{\partial^2}{\partial t^2} - \Delta\right)\left(\sigma_0 + \hat{\phi}\right) = \Delta\left(f_0 + \hat{f}\right). \tag{39}$$

Introducing the Maxwell potential definitions: $\varphi = \sigma_0 = \operatorname{div}\mathbf{u}_0$ and $\mathbf{A} = \hat{\phi} = \operatorname{rot}\mathbf{u}_\phi$ where they denote the irrotational scalar and solenoidal vector potentials, we get the four-potential and flux

$$A = \varphi + \mathbf{A} = \varphi + A_1 i + A_2 j + A_3 k, \tag{40}$$

$$\mu J = J_0 + \hat{J} = -\Delta f_0 - \Delta \hat{f} = -\Delta f. \tag{41}$$

The macroscopic version of the Maxwell equations is the formula (39) with the use of (40) and (41)

$$\left(-\dfrac{1}{c^2}\dfrac{\partial^2}{\partial t^2} + \Delta\right) A = \mu J. \tag{42}$$

The microscopic empty space version is as following. **C**onsider the empty crystal space (without the charged particles) and the irrotational deformations are negligible: $J = 0$. Consequently the relation (42) reduces to

$$\dfrac{1}{c^2}\dfrac{\partial^2 \mathbf{A}}{\partial t^2} + \operatorname{rot}\operatorname{rot}\mathbf{A} = 0. \tag{43}$$

We introduce definitions:

$$\mathbf{E} := -\dfrac{1}{c}\dfrac{\partial \mathbf{A}}{\partial t}, \tag{44}$$

$$\mathbf{H} := \operatorname{rot}\mathbf{A}. \tag{45}$$

Upon combining (43) - (45) and by taking the rotation in the definition (44): $\operatorname{rot}\left(c^{-1}\partial\mathbf{A}/\partial t = -\mathbf{E}\right)$ the Maxwell system for vacuum follows

$$\begin{cases} \dfrac{1}{c}\dfrac{\partial \mathbf{E}}{\partial t} - \operatorname{rot}\mathbf{H} = 0 \\ \dfrac{1}{c}\dfrac{\partial \mathbf{H}}{\partial t} + \operatorname{rot}\mathbf{E} = 0 \end{cases}. \tag{46}$$

## 2.4 The bosons, fermions, quarks and their q-potentials

### 2.4.1 The Quaternionic Propagators

The coupling of the transverse and the longitudinal waves takes place in the elementary cell of the Cauchy continuum, i.e., at the Planck scale. The quaternionic oscillator controls the acceleration of all the

$q$-potential components during the propagation, e.g., in the particle wave in $\Omega$: $\ddot{\sigma}_0, \ddot{\phi}_1, \ddot{\phi}_2, \ddot{\phi}_3$. The function $G_0 \in \mathbb{R}$ will be called the frequency of the oscillator. In the earlier papers, we disregarded that the twists $\phi_1, \phi_2$ and $\phi_3$ form the twist vector $\hat{\phi} = \phi_1 i + \phi_2 j + \phi_3 k$ [32] and are controlled by the oscillator $G_0$. Thus, the relation between the $q$-potential and its scalar component $\sigma_0$ will be corrected and consider the two $q$-potential constituents, $\sigma_0$ and $\hat{\phi}$ [32]:

$$\left\langle \frac{\partial^2 \sigma}{\partial t^2} \right\rangle = 2 \left\langle \frac{\partial^2 \sigma_0}{\partial t^2} \right\rangle = 8\pi^2 f_P f , \qquad (47)$$

and the frequency of the quaternionic oscillator equals

$$G_0(f) = 8\pi^2 f_P f . \qquad (48)$$

The particle wave frequency depends on the particle mass, $f = f(m)$, and follows from the $\mathbb{R}^1$ schema, e.g., see Fig. 1 in [16]. The sum of moments of all the Planck masses forming the particle wave in $\Omega$ (at the arbitrary time $t$ and solely due to the particle wave) equals the momentum of the particle $m$ itself. On the other hand, we may estimate the average momentum of the arbitrary single Planck mass $m_P$ in the elementary cell during the whole particle cycle: $T = f^{-1}$. The complete cycle implies that every Planck mass $m_P$ returns to its initial conditions: $\mathbf{u}_P(t) = \mathbf{u}_P(t+T)$ and $\dot{\mathbf{u}}_P(t) = \dot{\mathbf{u}}_P(t+T)$. The overall distance of the Planck mass during the wave cycle $T$ equals $2\pi l_P$. Thus, the average momentum of the Planck mass $\bar{p}(m_P)$ during the particle wave cycle equals

$$\bar{p}(m_P) = m_P \frac{2\pi l_P}{T} = 2\pi m_P l_P f . \qquad (49)$$

The momentum of the particle wave $m$ results from the particle wave propagation velocity:

$$p(m) = mc . \qquad (50)$$

The both moments must equal: $p(m) = \bar{p}(m_P)$, and the frequency of the particle wave becomes

$$f = \frac{mc}{2\pi m_P l_P} \frac{c}{c} = \frac{mc^2}{2\pi \hbar}, \quad \hbar = m_P c l_P , \qquad (51)$$

where upon using the NIST data [33] for the Planck natural units $m_P, l_P, t_P$ and the light velocity $c$, the constant $\hbar$ introduced in relation (51) equals the Planck constant [13]. Combining the relations (48), (51) and the definition $f_P = 1/t_P$, the overall frequency of the quaternionic oscillator when the particle mass is known equals

$$G_0(m) = 4\pi mc^2 / (\hbar t_P) . \qquad (52)$$

The oscillator might generate the lower frequencies $f$ of the particle wave and the families of propagators

$$G_n = \frac{1}{n} G_0(m) = \frac{1}{n} 4\pi mc^2 / (\hbar t_P), \quad \text{where } n = 1, 2, ... , \qquad (53)$$

where $n$ can be interpreted as the measure of the propagator volume, e.g., $l_n = n l_P$.

The quaternionic oscillator $G_0(m)$ controls four propagators:

- the scalar I (spin 0), $\qquad G_0(m) \sigma \cdot \sigma^*$,

- the scalar ½ (spin 1/2), $\quad G_0(m)\sigma_0,$
- the vectorial II (spin 1/2), $\quad G_0(m)\hat{\phi},$
- the quaternionic (spin 1/2), $\quad G_0(m)(\sigma_0+\hat{\phi}).$

The above propagators generate the particle wave and simultaneously, the particles produce different force fields that are represented by the Poisson equation

$$n c^2 \Delta \varphi + G_0(m) f = 0 \text{ where } \varphi \text{ and } f \text{ are the quaternion valued functions}. \qquad (54)$$

**Remark 3.** *Substituting $mc^2=E_0$ in (51), the Planck–Einstein relation follows: $E_0=hf$, where $h=2\pi\hbar$.*

### 2.4.2. Bosons

The family of the scalar second order PDE systems of the spin 0 particles results from schema in (16) and (53). In (55), we show the core set of the three second order PDE and its equivalent, the set of two second order equations: the particle wave and the force field produced by the particle. This schema will be used in the following sections.

$$\begin{cases}\left(\dfrac{\partial^2}{\partial t^2}-c^2\Delta\right)\hat{\phi}=0\\ \left(\dfrac{\partial^2}{\partial t^2}-3c^2\Delta\right)\sigma_0=0\\ nc^2\Delta\sigma_0+G_0(m)\sigma\cdot\sigma^*=0\end{cases}\Leftrightarrow\begin{cases}\left(\dfrac{\partial^2}{\partial t^2}-c^2\Delta\right)\tilde{\sigma}_n+2G_0(m)\sigma\cdot\sigma^*=0\\ \left((n-1)\dfrac{\partial^2}{\partial t^2}-(n-3)c^2\Delta\right)\sigma_0+2G_0(m)\sigma\cdot\sigma^*=0\end{cases}, \qquad (55)$$

where $\tilde{\sigma}_n=n\sigma_0+\hat{\phi}$ and $n$ denotes integer, $n\neq 0$. At $n=1$, the system (55) results in [15]:

$$\begin{cases}\left(\dfrac{\partial^2}{\partial t^2}-c^2\Delta\right)\hat{\phi}=0\\ \left(\dfrac{\partial^2}{\partial t^2}-3c^2\Delta\right)\sigma_0=0\\ c^2\Delta\sigma_0+G_0(m)\sigma\cdot\sigma^*=0\end{cases}\Leftrightarrow\begin{cases}\left(\dfrac{\partial^2}{\partial t^2}-c^2\Delta\right)\sigma+2G_0(m)\sigma\cdot\sigma^*=0\\ c^2\Delta\sigma_0+G_0(m)\sigma\cdot\sigma^*=0\end{cases}. \qquad (56)$$

The above two systems are equivalent and have five equations and five unknowns: $\sigma_0,\phi_1,\phi_2,\phi_3$ and $m$. If mass $m$ is unknown it may be treated as the parameter in the Poisson equation above. The equation (56) corresponds to the Klein-Gordon equation, i.e., the spin 0 boson particle.

The second order PDE systems (55) and (56) comply with the Cauchy equation of motion, i.e., by adding the Poisson and wave equations, the equation (16) results.

The Poisson equation in (56) describes the irrotational potential $\sigma_0$ of the deformation field

$$c^2\Delta\sigma_0=-G_0(m)\sigma\cdot\sigma^*=-4\pi\dfrac{mc^2}{\hbar t_P}\sigma\cdot\sigma^*. \qquad (57)$$

It can be expressed as a function of the particle mass density $\rho=m\sigma\cdot\sigma^*/l_P^3$:

$$c^2\Delta\sigma_0=-4\pi\rho\dfrac{l_P^3}{m_P t_P^2}=-4\pi\rho G. \qquad (58)$$

Using data in Table 1, the gravitational constant equals: $G = l_P^3/(t_P^2 m_P) = 6.674082 \cdot 10^{-11}\ [\mathrm{m^3 \cdot kg^{-1} \cdot s^{-2}}]$.
The particle mass center, equals its wave energy center. The "space-localized" particle is defined in the sense given by the Bodurov definition [34]:

"*A singularity-free multi-component function* $\sigma = (\sigma_0, \phi_1, \phi_2, \phi_3) \in \mathbb{Q}$ *of the space* $x = (x_1, x_2, x_3)$ *and time t variables will be called space-localized if* $\|\sigma(t,x)\| \to 0$ *sufficiently fast when* $\|x\| \to \infty$, *so that its Hermitean norm*

$$\|\sigma\|^2 = \langle \sigma, \sigma^* \rangle = \int_\Omega \left( \sigma_0^2 + \sum_{l=1}^{3} \phi_l \cdot \phi_l^* \right) dx = \int_\Omega \sigma \cdot \sigma^* dx < \infty \tag{59}$$

*remains finite for all time.*"

### 2.4.3. The particles formed by the odd number of quarks

The strong coupling is considered in the following sections: $n = 1$ in the relation (53).

We begin with the vectorial potential, where the term $G_0(m)\hat{\phi}$ fixes the density of the rate of twist change and is called vectorial propagator

$$\begin{cases} \left( \dfrac{\partial^2}{\partial t^2} - c^2 \Delta \right) \hat{\phi} = 0 \\ \left( \dfrac{\partial^2}{\partial t^2} - 3c^2 \Delta \right) \sigma_0 = 0 \\ -c^2 \Delta \hat{\phi} + G_0(m) \hat{\phi} = 0 \end{cases} \Leftrightarrow \begin{cases} \left( \dfrac{\partial^2}{\partial t^2} - 3c^2 \Delta \right) \sigma + 2 G_0(m) \hat{\phi} = 0 \\ -c^2 \Delta \hat{\phi} + G_0(m) \hat{\phi} = 0 \end{cases}. \tag{60}$$

Upon the rearrangement, the particle wave (electron) and the vectorial Poisson equations are evident. The adding equations in the system (60) shows that it complies with the Cauchy equation of motion (16):

$$\begin{cases} \left( \dfrac{\partial^2}{\partial t^2} - 3c^2 \Delta \right) \sigma + 2 G_0(m) \hat{\phi} = 0, \\ -c^2 \Delta \hat{\phi} + G_0(m) \hat{\phi} = 0, \end{cases} \Rightarrow \left( \dfrac{\partial^2}{\partial t^2} - c^2 \Delta \right) \sigma - 2c^2 \Delta \sigma_0 = 0. \tag{61}$$

Note that the wave propagation velocity in system (60) equals the velocity of longitudinal waves in the Cauchy continuum: $c_L = \sqrt{3}\, c$ [13]. The vectorial Poisson equation in (60) confirms that it is the second order PDE system for electron.

In the quaternion propagator, $G_0(m)(\sigma_0 + \hat{\phi})$, the vectorial, $G_0(m)\hat{\phi}$, and scalar, $G_0(m)\sigma_0$, propagators are "merged" and form the strongly coupled system. The rearrangement of the system (62) is shown below and displays different forms of the second order PDE systems:

$$\begin{cases} \left( \dfrac{\partial^2}{\partial t^2} - c^2 \Delta \right) \hat{\phi} = 0 \\ \left( \dfrac{\partial^2}{\partial t^2} - 3c^2 \Delta \right) \sigma_0 = 0 \\ -c^2 \Delta \hat{\phi} + G_0(m)\hat{\phi} = 0 \\ c^2 \Delta \sigma_0 + G_0(m)\sigma_0 = 0 \end{cases} \Leftrightarrow \begin{cases} \left( \dfrac{\partial^2}{\partial t^2} - c^2 \Delta \right) \hat{\phi} = 0 \\ \left( \dfrac{\partial^2}{\partial t^2} - 3c^2 \Delta \right) \sigma_0 = 0 \\ c^2 \Delta (\sigma_0 - \hat{\phi}) + G_0(m)(\sigma_0 + \hat{\phi}) = 0 \end{cases} \Leftrightarrow \begin{cases} \left( \dfrac{\partial^2}{\partial t^2} - 2c^2 \Delta \right) \sigma + G_0(m)(\sigma_0 + \hat{\phi}) = 0 \\ c^2 \Delta \sigma^* + G_0(m) \sigma = 0 \end{cases}. \tag{62}$$

The comparison of the scalar, vectorial and quaternionic propagators shows that the q-propagator offers the strongest coupling, Eq. (62). The quaternionic Poisson equation in (62) reveals that it is the second order PDE system for proton. The sum of equations in (62) shows that system complies with the Cauchy equation (16):

$$\begin{cases}\left(\dfrac{\partial^2}{\partial t^2}-2c^2\Delta\right)\sigma+G_0(m)\left(\sigma_0+\hat{\phi}\right)=0 \\ c^2\Delta\left(\sigma_0-\hat{\phi}\right)+G_0(m)\left(\sigma_0+\hat{\phi}\right)=0\end{cases} \Rightarrow \left(\dfrac{\partial^2}{\partial t^2}-c^2\Delta\right)\sigma-2c^2\Delta\sigma_0=0. \tag{63}$$

Note that the propagation velocity in system (62) exceeds the transverse wave velocity: $c'=\sqrt{2}\,c$.

**2.4.4 The quarks.** The comparison of the systems (56), (60) and (62) allows postulating the second order PDE for the quarks from the *up* and *down* groups. Explicitly, the second order system of the *u* quark from the *up* group equals:

$$\begin{cases}\left(\dfrac{1}{3}\dfrac{\partial^2}{\partial t^2}-c^2\Delta\right)\sigma+\dfrac{2}{3}G_0(m)\hat{\phi}=0 \\ -c^2\dfrac{2}{3}\Delta\hat{\phi}-\dfrac{2}{3}G_0(m)\hat{\phi}=0\end{cases} \tag{64}$$

and the system of the *d* quark from the *down* group

$$\begin{cases}\dfrac{1}{3}\dfrac{\partial^2\sigma}{\partial t^2}+G_0(m)\sigma_0-\dfrac{1}{3}G_0(m)\hat{\phi}=0 \\ c^2\Delta\left(\sigma_0+\dfrac{1}{3}\hat{\phi}\right)-G_0(m)\left(\sigma_0-\dfrac{1}{3}\hat{\phi}\right)=0\end{cases}. \tag{65}$$

The sum of equations in the above quark systems (64) and (65) does not comply with the Cauchy equation of motion (16) and may indicate their short lifetime. The terms $\dfrac{2}{3}G_0(m)\hat{\phi}$ and $-\dfrac{1}{3}G_0(m)\hat{\phi}$ in the systems (64) and (65) respectively, are related to the charge, Table 1.

**Table 1.** The basic properties of the quarks in baryons.

| Group | Quarks | Charge | Spin |
|---|---|---|---|
| *up* | u,c,t | 2/3 | ½ |
| *down* | d,s,b | -1/3 | ½ |

There are two groups of hadrons: baryons (containing three quarks or three antiquarks); and mesons (containing a quark and an antiquark). In the following we show that systems (60) - (65) comply with the experimental findings shown in Table 1.

Proton is formed be the two up and the single down quarks: $d-u-u$. Thus by computing the sum of two systems (64) and one system (65) we may expect the proton, the system (62):

$$\begin{cases}\dfrac{1}{3}\dfrac{\partial^2\sigma}{\partial t^2}+G_0(m)\sigma_0-\dfrac{1}{3}G_0(m)\hat{\phi}=0 \\ c^2\Delta\left(\sigma_0+\dfrac{1}{3}\hat{\phi}\right)+G_0(m)\left(\sigma_0-\dfrac{1}{3}\hat{\phi}\right)=0\end{cases} + 2\begin{cases}\left(\dfrac{1}{3}\dfrac{\partial^2}{\partial t^2}-c^2\Delta\right)\sigma+\dfrac{2}{3}G_0(m)\hat{\phi}=0 \\ -c^2\dfrac{2}{3}\Delta\hat{\phi}+\dfrac{2}{3}G_0(m)\hat{\phi}=0\end{cases} \tag{66}$$

and the result is identical with equations (62)

$$\begin{cases}\left(\dfrac{\partial^2}{\partial t^2}-2c^2\Delta\right)\sigma+G_0(m)\left(\sigma_0+\hat{\phi}\right)=0 \\ c^2\Delta\left(\sigma_0-\hat{\phi}\right)+G_0(m)\left(\sigma_0+\hat{\phi}\right)=0\end{cases}. \tag{67}$$

Neutron is formed by the one up and the two down quarks: $d-d-u$

$$2\times\begin{cases}\dfrac{1}{3}\dfrac{\partial^2\sigma}{\partial t^2}+G_0(m)\sigma_0-\dfrac{1}{3}G_0(m)\hat{\phi}=0\\ c^2\Delta\left(\sigma_0+\dfrac{1}{3}\hat{\phi}\right)+G_0(m)\left(\sigma_0-\dfrac{1}{3}\hat{\phi}\right)=0\end{cases}+\begin{cases}\left(\dfrac{1}{3}\dfrac{\partial^2}{\partial t^2}-c^2\Delta\right)\sigma+\dfrac{2}{3}G_0(m)\hat{\phi}=0\\ -c^2\dfrac{2}{3}\Delta\hat{\phi}+\dfrac{2}{3}G_0(m)\hat{\phi}=0\end{cases} \quad (68)$$

and the result is in agreement with neutron system (56)

$$\begin{cases}\dfrac{\partial^2\sigma}{\partial t^2}-c^2\Delta\sigma+2G_0(m)\sigma_0=0\\ c^2\Delta\sigma_0+G_0(m)\sigma_0=0\end{cases}. \quad (69)$$

The systems (60), (67) and (69) represent coupled second order PDE's and show the different coupling strengths. The strongest coupling of the proton is related to its enormously long lifetime, Equation (67).

## 2.5. The Quaternionic Schrödinger Equation

The vectorial Poisson equation indicates that it's the second order PDE system for electron. We will apply the schema in the system (60) in the integral form of the energy conservation, in Equation (21). We treat the wave as a particle in an arbitrary volume $\Omega$ [15]. The energy per mass unit, $e$

$$e=\frac{1}{2}\hat{u}\cdot\hat{u}^*+\frac{3}{2}c^2\sigma\cdot\sigma^*-c^2\hat{\phi}\cdot\hat{\phi}^* \quad (70)$$

in the volume occupied by the particle wave defines its overall energy

$$E_O=E_P+E_V=\int_\Omega\rho_P e\,dx, \quad (71)$$

where $E_p$ and $E_V$ are energies of the particle and of its force field respectively, $\rho_P$ is the Planck mass density.

The first step in deriving the Schrödinger equation is the choice of the symmetrization scheme for the particle energy, $E_p$. Equation (71) can be written in the equivalent form following schema in system (60). We separate the $E_p$ and $E_V$ terms in integral formula

$$E_P+E_V=\rho_P\int_\Omega\left(\frac{1}{2}\hat{u}\cdot\hat{u}^*+\frac{3}{2}c^2\sigma\cdot\sigma^*-c^2\hat{\phi}\cdot\hat{\phi}^*\right)dx\Leftarrow\begin{cases}E_P=\dfrac{1}{2}\rho_P\int_\Omega\left(\hat{u}\cdot\hat{u}^*+3c^2\sigma\cdot\sigma^*\right)dx\\ E_V=\rho_P\int_\Omega\left(-c^2\hat{\phi}\cdot\hat{\phi}^*\right)dx\end{cases}. \quad (72)$$

The mass of the particle $m=E_p/c^2$ follows from the particle wave energy in (72)

$$m=\frac{1}{2}\rho_P\int_\Omega\left(3\sigma\cdot\sigma^*+\frac{\hat{u}\cdot\hat{u}^*}{c^2}\right)dx. \quad (73)$$

The terms $3\sigma\cdot\sigma^*$ and $\hat{u}\cdot\hat{u}^*/c^2$ oscillate and depend on the time and position. The symmetry in (73) allows normalizing the deformation and mass velocity with respect to the overall particle mass

$$\int_\Omega\frac{3\rho_P}{m}\sigma\cdot\sigma^*dx=\int_\Omega\psi\cdot\psi^*dx=1,\quad\text{where }\psi=\sqrt{\frac{3\rho_P}{m}}\,\sigma,$$

$$\int_\Omega\frac{\rho_P}{mc^2}\hat{u}\cdot\hat{u}^*dx=\int_\Omega\psi\cdot\psi^*dx=1,\quad\text{where }\psi=\sqrt{\frac{\rho_P}{m}}\,\frac{\hat{u}}{c}. \quad (74)$$

The quaternionic particle mass density $\psi$ can be called the quaternionic probability because the relation $\int_\Omega\psi\cdot\psi^*dx=1$ in (74) is satisfied. Obviously, terms $\psi=\sqrt{3\rho_P/m}\,\sigma(t,x)$ and $\psi\cdot\psi^*$, vary in time.

We analyze the evolution of the wave as in relations (73) and (74) in the time-invariant potential field, e.g., the particle wave in the field generated by other particles. The overall particle energy is now a sum of the ground and excess energy $Q$,

$$E = E_p + Q = \int_\Omega \left( \frac{3}{2} \rho_P c^2 \sigma \cdot \sigma^* + \frac{1}{2} \rho_P \hat{u} \cdot \hat{u}^* + V(x) \psi \cdot \psi^* \right) dx . \tag{75}$$

We consider the low excess energies, and the impact of $Q$ on the overall particle mass in (74) is marginal. Thus, the relation (75) becomes

$$E = E_p + Q = \int_\Omega \left( \frac{1}{2} m c^2 \psi \cdot \psi^* + \frac{1}{2} \rho_P \hat{u} \cdot \hat{u}^* + V(x) \psi \cdot \psi^* \right) dx$$
$$= \frac{1}{2} m c^2 + \int_\Omega \left( \frac{1}{2} \rho_P \hat{u} \cdot \hat{u}^* + V(x) \psi \cdot \psi^* \right) dx . \tag{76}$$

Both the $E_p$ and $m$ are constant; thus, it is enough to minimize the relation

$$Q = \int_\Omega \left( \frac{1}{2} \rho_P \hat{u} \cdot \hat{u}^* + V(x) \psi \cdot \psi^* \right) dx . \tag{77}$$

The above relation contains two unknowns: $\hat{u} = \partial u / \partial t$ and $\psi$. By relating the local lattice velocity $\hat{u}$ to the force, specifically to the normalized Cauchy–Riemann derivative of the deformation $l_P D\sigma$, one gets

$$\hat{u} = \frac{\hat{p}}{m} = -\frac{\hbar}{m} D\sigma . \tag{78}$$

By introducing (78) and the normalization (74), the relation (77) becomes the functional

$$Q[\psi] = \int_\Omega \left( \frac{\hbar^2}{2m} (D\psi) \cdot (D\psi)^* + V(x) \psi \cdot \psi^* \right) dx . \tag{79}$$

The functional $Q[\psi]$, Eq. (79), was minimized with respect to a quaternion function, such that $\psi$ satisfies the normalization introduced in the relation . One may follow the schema used in [15]. In simple terms, we seek a differential equation that has to be satisfied by the $\psi$ function to minimize the energies allowed by (79). Given the functional (79), the conditional extreme is found using the Lagrange coefficients method and the Du Bois Reymond variational lemma [35]. In such a case, $\psi$ satisfies the time-invariant Schrödinger equation satisfied by the particle wave in the ground state of the energy $E$ [15]:

$$-\frac{\hbar^2}{2m} \Delta \psi + V(x) \psi = \lambda \psi , \tag{80}$$

where a constant factor on the right-hand side can be considered as extra energy of the particle in the presence of the field $V = V(x)$. For $E = \lambda$, Equation (80) is clearly the time-independent Schrödinger equation satisfied by the particle in the ground state of the energy $E$,

$$-\frac{\hbar^2}{2m} \Delta \psi + V(x) \psi = E\psi . \tag{81}$$

### 2.6. The First Order PDE in the Cauchy Continuum

The operator quantum mechanics base on the complex number algebra and the matrix algebra. The canonical quantization starts from the classical mechanics and assumes that the point particle is described by a "probabilistic wave function". Dirac applied complex combinations of the displacements and velocities in the linear problem of secondary quantization [36] and replaced the second order Klein–Gordon equation by an array of first order equations. He recognized the problem of medium for the transmission of waves:

*"It is necessary to set up an action principle and to get a Hamiltonian formulation of the equations suitable for quantization purposes, and for this the aether velocity is required"* [37].

In this section we follow the Dirac concept. We derive the formulae basing on the aether concept. Explicitly, the Cauchy continuum and the quaternionic oscillator $G_\lambda(m)$ for the first order PDE and the separated Planck time scale. The second order particle wave equation, e.g., in the electron PDE system (61), contains two parts:

$$\left(\frac{\partial^2}{\partial t^2} - c_L^2 \Delta\right)\sigma \quad + \quad 2G_0(m)\hat{\phi} \quad = \quad 0$$

$$\begin{bmatrix} \text{second order wave term } \sigma^\mu\sigma_\mu \text{: variable} \\ \sigma \text{ and constant wave velocity } c_L = \sqrt{3}\,c \end{bmatrix} + \begin{bmatrix} \text{Propagator with oscillator } G_0(m) \\ \text{that runs at two frequencies} \end{bmatrix} = 0 \quad . \tag{82}$$

We will comply with above schema for the first order PDE:

$$\left(\frac{\partial}{\partial t} - c_L D\right)\frac{\hat{u}}{c_L} \quad + \quad 2G_\lambda(m)\frac{\hat{u}}{c_L} \quad = \quad 0$$

$$\begin{bmatrix} \text{First order wave term } \sigma_\mu \text{: variable} \\ \hat{u}/c_L \text{, wave velocity } c_L = \sqrt{3}\,c \end{bmatrix} + \begin{bmatrix} \text{Propagator with oscillator } G_\lambda(m) \\ \text{that runs at particle wave frequency} \end{bmatrix} = 0 \quad . \tag{83}$$

### 2.6.1. The first order wave term.

We consider the system (60) and the relation between the wave velocity and the Cauchy–Riemann derivative Equation (78): $D\sigma = -\frac{m}{\hbar}\hat{u}$. The expression for the overall particle energy, Equation (72), implies:

- the deformation velocity as the alternative variable:

$$\frac{\hat{u}}{c_L} = -\frac{\hbar}{mc_L}D\sigma, \tag{84}$$

- the longitudinal wave velocity as the wave propagation velocity:

$$c_L = \sqrt{3}\,c. \tag{85}$$

The motionless particle is considered, thus its wave is at a steady state. The second order time derivative of the q-potential in (83) we express as follows:

$$\frac{\partial^2 \sigma}{\partial t^2} = \frac{\partial}{\partial t}\left(\frac{\partial \sigma}{\partial t}\right). \tag{86}$$

The term in the bracket on the right-hand side is the rate of the q-potential changes. We want to express this term by the new variable and separate the time scales. The rate of changes of the deformation potential $\partial\sigma/\partial t$ is due to the wave propagation within the particle space. The propagation process must follow the extremum principle, i.e., it is the brachistochrone problem [38]. The good example of "local principle" approximation is by Derbes [39].

We know that the wave path fulfills the extremum principle, i.e., the wave path follows its unique trajectory given by the Cauchy–Riemann derivative, $D\sigma$. The trajectory which has the minimum property globally in the whole volume $\Omega$ occupied by the particle must have the same property locally. This path grants the shortest possible travelling time for the waves identified in QQM. Consequently from (84) - (86) we postulate the following:

$$\begin{cases} \dfrac{\partial \sigma}{\partial t} = \dfrac{\partial \mathbf{u}}{\partial t}\left(\dfrac{\partial \sigma}{\partial \mathbf{u}}\right) \\ \dfrac{\partial \mathbf{u}}{\partial t} = c_L \\ \dfrac{\partial \sigma}{\partial \mathbf{u}} = \mathrm{D}\sigma = -\dfrac{m}{\hbar}\hat{u} \end{cases} \Rightarrow \quad \dfrac{\partial \sigma}{\partial t} = c_L \mathrm{D}\sigma = \dfrac{mc_L}{\hbar}\hat{u}\ . \tag{87}$$

From the relation (84) we get

$$\mathrm{D}\sigma = -\dfrac{m}{\hbar}\hat{u} \quad \Rightarrow \quad \Delta\sigma = -\mathrm{DD}\sigma = \dfrac{m}{\hbar}\mathrm{D}\hat{u}. \tag{88}$$

Combining the relations (87) and (88), we get the first order particle wave term consistent with the second order formula (82):

$$\dfrac{\partial^2 \sigma}{\partial t^2} - c_L^2 \Delta\sigma \Leftrightarrow \dfrac{m}{\hbar}\dfrac{\partial \hat{u}}{\partial t} - \dfrac{mc_L}{\hbar}\mathrm{D}\hat{u} = \dfrac{mc_L}{\hbar}\left(\dfrac{\partial}{\partial t} - c_L \mathrm{D}\right)\dfrac{\hat{u}}{c_L}. \tag{89}$$

Thus, the first order particle wave term in (83) equals:

$$\left(\dfrac{\partial}{\partial t} - c_L \mathrm{D}\right)\dfrac{\hat{u}}{c_L} = 0. \tag{90}$$

**The first order quaternionic oscillator.** The frequency of the second order quaternionic oscillator results from two time scales in PK-C: $G_0(f) = 8\pi^2 f_P f$, . We consider the macro scale only and first order PDE equation thus, by eliminating the Planck frequency from the relation (48), results in the frequency formula of the first order quaternionic oscillator when the particle mass is known:

$$G_\lambda(m) = 4\pi f = 2\dfrac{mc_L^2}{\hbar} = 6\dfrac{m}{m_P t_P}. \tag{91}$$

By introducing the relations (90) and (91) in the schema (83), the first order PDE for electron equals

$$\left(\dfrac{\partial}{\partial t} - c_L \mathrm{D}\right)\dfrac{\hat{u}}{c_L} - 6\dfrac{m}{m_P t_P}\hat{\phi} = 0. \tag{92}$$

By substituting (84)

$$\left(\dfrac{1}{c_L}\dfrac{\partial}{\partial t} - \mathrm{D}\right)\dfrac{\hbar}{m}\mathrm{D}\sigma + \dfrac{6m}{m_P t_P}\hat{\phi} = 0. \tag{93}$$

**The Electron Spin.** The energy relations (72) are symmetrical and, in the case of the electron:

$$\begin{cases} E_{\substack{electron \\ particle}} = \dfrac{1}{2}\rho_P \int_\Omega \left(\hat{u}\cdot\hat{u}^* + c_L^2 \sigma\cdot\sigma^*\right)\mathrm{d}x \\ E_{\substack{electron \\ potential\ field}} = \rho_P \int_\Omega \left(-c^2 \hat{\phi}\cdot\hat{\phi}^*\right)\mathrm{d}x \end{cases} . \tag{94}$$

A particle is stable and its energy must be conserved. Thus, it is justified to assume that the constraint $\mathrm{div}\,\hat{\phi} = 0$ holds for the completed particle cycle. In the static particle we postulate zero dissipation of the twist energy: $\mathrm{div}\,\hat{\phi} = 0$. It implies the necessity of the spin, $\hat{S}$, the process that will provide the energy conservation:

$$\mathrm{div}\,\hat{\phi} = \mathrm{div}\left(\hat{\phi}' + \hat{S}\right) = 0. \tag{95}$$

The equipartition of energy between the twists, $|\hat{\phi}'| = |\hat{S}|$, implies the equipartition of moments: $|\hat{\phi}'| = |\hat{S}|$.

Thus, the overall momentum per mass unit equals: $2|\hat{S}| = \alpha$.

## 3. Mathematical model of the electron

We consider the quaternionic system of equations shown in Equation (61)

$$\begin{cases} \dfrac{\partial^2 \sigma}{\partial t^2} - 3c^2 \Delta \sigma + 2G_0(m)\hat{\phi} = 0 \\ c^2 \Delta \hat{\phi} - G_0(m)\hat{\phi} = 0 \end{cases}. \tag{96}$$

The system (96) is equivalent to the real hyperbolic-elliptic system

$$\begin{cases} \dfrac{\partial^2 \sigma_0}{\partial t^2} - 3c^2 \Delta \sigma_0 = 0 \\ \dfrac{\partial^2 \hat{\phi}}{\partial t^2} - 3c^2 \Delta \hat{\phi} + 2G_0(m)\hat{\phi} = 0 \\ c^2 \Delta \hat{\phi} - G_0(m)\hat{\phi} = 0 \end{cases} \tag{97}$$

that can be written in the elegant mathematical form

$$\begin{cases} \dfrac{\partial^2 \sigma_0}{\partial t^2} - 3c^2 \Delta \sigma_0 = 0 \\ \dfrac{\partial^2 \hat{\phi}}{\partial t^2} - c^2 \Delta \hat{\phi} = 0 \\ c^2 \Delta \hat{\phi} - G_0(m)\hat{\phi} = 0 \end{cases}. \tag{98}$$

Each of the three equivalent systems (96) - (98) obeys the constraints

$$\begin{cases} \operatorname{div} \hat{\phi} = 0 \\ \phi_1 = 2\phi_2 \end{cases}. \tag{99}$$

The energy must be conserved. Thus the flux, Eq. (31), through the boundary must be zero and we consider the boundary condition in the form

$$\left(c^2 \hat{\phi} \times \hat{\dot{u}} - 3c^2 \sigma_0 \hat{\dot{u}}\right) \circ \hat{n} = 0, \tag{100}$$

where

$$\dot{u} = -\dfrac{\hbar}{m}\left(\nabla \sigma_0 + \operatorname{rot} \hat{\phi}\right). \tag{101}$$

Combining (100) and (101) results in

$$\left(-\hat{\phi} \times \nabla \sigma_0 - \hat{\phi} \times \operatorname{rot} \hat{\phi} + 3\sigma_0 \nabla \sigma_0 + 3\sigma_0 \operatorname{rot} \hat{\phi}\right) \circ \hat{n} = 0. \tag{102}$$

Electron does not generate the scalar field and we add the condition

$$\nabla \sigma_0 \circ \hat{n} = 0. \tag{103}$$

Consequently the condition (102) becomes

$$\left(\hat{\phi} \times \operatorname{rot} \hat{\phi} - 3\sigma_0 \operatorname{rot} \hat{\phi}\right) \circ \hat{n} = 0. \tag{104}$$

The twist $\hat{\phi}$ and compression $\sigma_0$ in the above differential problem (98) are coupled only at the boundary, Eq. (104). In particular, $\sigma_0$ affects $\hat{\phi}$ but not vice versa. So, one may formulate two initial-boundary value problems.

### 3.1. Initial-boundary value problem for compression $\sigma_0$

The hyperbolic equation

$$\frac{\partial^2 \sigma_0}{\partial t^2} - 3c^2 \Delta \sigma_0 = 0 \tag{105}$$

with the Neumann boundary condition

$$\nabla \sigma_0 \circ \hat{n} = 0 \tag{106}$$

and initial conditions

$$\begin{cases} \sigma_0(0, x) = \sigma_{01}(x) \\ \dfrac{\partial \sigma_0}{\partial t}(0, x) = \sigma_{02}(x) \end{cases}, \tag{107}$$

where the functions $\sigma_{01}$ and $\sigma_{02}$ are given.

### 3.2. Initial-boundary value problem for twist $\hat{\phi}$

The hyperbolic-elliptic system

$$\begin{cases} \dfrac{\partial^2 \hat{\phi}}{\partial t^2} - c^2 \Delta \hat{\phi} = 0 \\ c^2 \Delta \hat{\phi} - G_0(m) \hat{\phi} = 0 \\ \operatorname{div} \hat{\phi} = 0 \\ \phi_1 = 2\phi_2 \end{cases} \tag{108}$$

with the boundary condition

$$\left(\hat{\phi} \times \operatorname{rot} \hat{\phi} - 3\sigma_0 \operatorname{rot} \hat{\phi}\right) \circ \hat{n} = 0 \tag{109}$$

and the initial conditions

$$\begin{cases} \hat{\phi}(0, x) = \hat{\phi}_{01}(x) \\ \dfrac{\partial \hat{\phi}}{\partial t}(0, x) = \hat{\phi}_{02}(x) \end{cases}, \tag{110}$$

where the functions $\hat{\phi}_{01}$ and $\hat{\phi}_{02}$ are given.

The numerical solution of the model will be presented in the next paper.

## 4. Conclusions

The presented results are based on the ontological model of the QQM and QFT, i.e., on the Cauchy continuum and the Planck unit cell concepts. The major progress is due to the symmetrization of quaternion relations. Explicitly, due to the postulate of the quaternion velocity. It allows considering the momentum of the expanding Cauchy continuum and is the apparent result of the scalar potential $\sigma_0(t,x)$ of the expansion/compression. The key new results are listed below:

- The quaternionic $G_0(m)(\sigma_0 + \hat{\phi})$, vectorial $G_0(m)\hat{\phi}$ and scalar: $G_0(m)\sigma_0$, propagators are postulated and used to generate the second order PDE systems for the proton, electron and neutron.
- The scrupulous assessment of the second order PDE systems allows postulating the two second order PDE systems for the *u* and *d* quarks from the *up* and *down* groups.
- It is shown that both the proton and the neutron obey experimental findings and are formed by three quarks. Namely, the proton and neutron are formed by *d-u-u* and *d-d-u* complexes, respectively. All the above PDE systems comply with the Cauchy equation of motion (16) and can be considered as stable particles.
- The *u* and *d* quark systems do not comply with the Cauchy equation of motion. Also experimental efforts to find the individual quarks were without success. Observed were the bound states of the three quarks – the baryons and a quark and an antiquark – the mesons. Wilczek calls it the phenomenological paradox: *"Quarks are Born Free, but Everywhere They are in Chains"* [40]. The inconsistency of the quarks PDES with the Cauchy equation of motion elucidates the observed *Quarks Chains.*

The principle of special relativity is explained by this model and may well be applied for all practical purposes. The matter almost completely is built up by electromagnetic forces, so we must expect that both Lorentz contraction and time dilatations will be exactly as predicted by the Lorenz transformations. In the light of this prediction, it is not strange that the experiment by Michelson and Morley and by many others, did not reveal the speed of earth through the spatial continuum, which at that time was called the aether.

The results indicate the following targets for an immediate future:

- The particles and quarks in the case of higher coupling coefficients: $n > |1|$.
- The ratios between the constants for the different force fields.
- The rigorous derivation of the first order PDE basing on the extremum principle.
- The multivalued coordinate transformation to determine the properties of space with curvature and torsion produced by the second order PDE systems representing the QFT [41].

**Funding:** This research received no external funding.

**Acknowledgements:** The ideas reported here were developed during several discussions with Chantal Roth. Her criticism and suggestions were essential in the present QQM formulation. We owe her our profound thanks.

# Appendix A
*Abbreviations*

| | |
|---|---|
| PDE | partial differential equation |
| $\mathbb{Q}$ | quaternion algebra |
| QQM | quaternion quantum mechanics |
| QFT | quaternion field theory |
| **T** | deformation tensor |
| $\lambda, \mu$ | Lamé coefficients; |
| $\sigma'$ | stress tensors |
| $\rho_E$ | density of the deformation energy |
| $\mathbf{u}(u_1, u_2, u_3)$ | displacement in $\mathbf{R}^3$ |
| $\sigma(\sigma_0, \phi_1, \phi_2, \phi_3)$ | *q-potential* in $\mathbf{R}^4$, the quaternion deformation potential |
| $\sigma^* \cdot \sigma$ | strain energy density |
| $G_0$ | frequency of the quaternionic oscillator |
| $G_0 \sigma^* \cdot \sigma$ | quaternionic scalar propagator I |
| $G_0 \sigma$ | quaternionic scalar propagator II |
| $G_0 \hat{\phi}$ | quaternionic vectorial propagator |
| $G_0(m)(\sigma \cdot \sigma^* + \hat{\phi})$ | quaternionic q-potential propagator |
| $\psi = \sigma \sqrt{\rho_P / m}$ | quaternionic particle density, i.e., the particle wave function |
| $\psi \cdot \psi^*$ | probability, i.e., the normalized particle mass density |
| $N$ | coupling coefficient in the oscillator |
| $l_P$ | Planck length |
| $f_P = 1/t_P$ | Planck frequency, inverse of the Planck time |
| $m_P$ | Planck mass |
| $c = l_P / t_P$ | transverse wave velocity in elastic continuum |
| $c_L = \sqrt{3}\, c$ | longitudinal wave velocity in elastic continuum |
| $\rho_P = 4 m_P / l_P^3$ | Planck density, i.e., the mass density of the PK-C |
| $\rho$ | mass density of the particle $\rho = \rho_E / c^2$, as the equivalent of the energy density $\rho_E$ in the PK-C |
| $\hbar$ | Planck constant in terms of angular frequency |

| | | |
|---|---|---|
| $h$ | | Planck constant, $h = 2\pi\hbar$ |
| $m$ | | equivalent mass of the wave, i.e., mass of the particle |
| $\lambda$ | | length of the particle wave |
| $f$ | | frequency of the particle wave |